\documentstyle[aps,prb,psfig]{revtex}
\begin{document} 
%\draft
%\preprint{}               
\title{Resonant versus anti-resonant tunneling at carbon nanotube A-B-A heterostructures}
\author{N.Mingo, Liu Yang, Jie Han \cite{byline} and M.P.Anantram}
\address{ NASA-Ames Research Center, Mail Stop T27A-1, Moffett Field, CA 94035-1000}
%\date{\today}
\maketitle
\begin{abstract}                % DON'T CHANGE THIS LINE
Narrow antiresonances going to zero transmission are found to occur for general (2n,0)(n,n)(2n,0) carbon nanotube
heterostructures, whereas the complementary configuration, (n,n)(2n,0)(n,n), displays simple
resonant tunneling behaviour. We compute examples for different cases, and give a simple
explanation for the appearance of antiresonances in one case but not in the other. Conditions
and ranges for the occurence of these different behaviors are stated. The phenomenon of
anti-resonant tunneling, which has passed unnoticed in previous studies of nanotube heterostructures,
adds up to the rich set of behaviors available to nanotube based quantum effect devices.
\\
\\
\bf{Published in Physica Status Solidi (b) 226, No.1, 79-85 (2001)}\\
\\
\end{abstract}

%\narrowtext
\begin{text}
Rapid progress in carbon nanotube electronic devices has been made recently \cite{Saito,Odom}.  Nanotube transistors \cite{Transistor} and diodes \cite{Diode} have been experimentally verified. Doping of different sections of the tube has been proposed for the fabrication of devices, \cite{Esfarjani}. Nevertheless, one of the most interesting
ways of producing a nanotube device consists in forming heterojunctions \cite{Chico,Treboux,Han} where different chiral tubes are joined by carbon pentagons and heptagons. This
allows to fully exploit the metallic (m) and semiconducting (s) characters
of nanotubes on a pure carbon molecule without
doping. The main question is to find out the electronic behaviour of these nanotube heterostructures. Here we address the problem of three component heterostructures of type A-B-A, where A and B stand for different chirality of the tube.

In solid state nanoelectronic quantum devices, an A-B-A device behaves as a single tunneling barrier 
or as a quantum well, depending on the relative energies of the conduction band edges of A and B, and 
one needs an A-B-A'-B-A double barrier heterostructure in order to obtain resonant tunneling. 
On the contrary, we will show that nanotube ABA junctions can display simultaneously bound
states, single barrier tunneling, and double barrier resonant
tunneling. Furthermore, several kinds of resonant tunneling can occur, more complex than the 'Fabry-Perot'-like behaviour, depending on the 
chirality of the A and B sections and junction geometry. The conditions for each kind of tunneling to occur can be established. This suggests 
nanotube ABA junctions as potential quantum effect devices. A recent proposal showing the robustness of a 
(5,5)(6,4)(5,5) junction as quantum device has been published \cite{Hyldgaard}. However, the possibility of
anti-resonant tunneling in nanotube heterostructures has largely been unnoticed in previous studies.

In this work, straight ABA heterostructures were constructed \cite{bend}
by joining two straight AB junctions. Here, A and B stand for different orientations of the 
graphene hexagons with respect to the nanotube translation axis. The orientation is usually 
specified by the 2-D chiral vector, defining the section of the nanotube perpendicular to its
translation axis on a 2-D honeycomb lattice \cite{Saito}. In terms of the honeycomb basis vectors,
the chiral vector has the form $(n,m)$, where $n$ and $m$ are integers. Tube segments with
$(n,0)$ or $(n,n)$ chirality are called zigzag and armchair tubes respectively (see fig.1a).
The geometries considered were further relaxed by a molecular dynamic
simulation approach, as reported previously \cite{Han}.  The calculation of their conductance curves
were then carried out in a Linear Combination of Atomic Orbitals framework 
\cite{Yang}.
The conductance across any interface dividing the system can be computed as \cite{Mingo}

\begin{eqnarray}
\sigma= {\scriptstyle{8e \over h}}
Tr[D_{11} \rho_{11} T_{12}
D_{22}^\dagger \rho_{21}^\dagger T_{22}^\dagger ].
\end{eqnarray}

with the denominators D given by

\begin{eqnarray}
D_{11(22)}= 
[I - T_{12(21)} G_{22(11)} T_{21(12)} G_{11(22)}]^{-1}.
\end{eqnarray}

$G$ is the retarded Green function of the decoupled systems, and $T$ is the hopping matrix joining the two parts. $\rho$ corresponds to $Im[G]/\pi$, and the self energies of the semi-infinite electrodes are 
calculated as described in \cite{Decimation}. To calculate the Green function we took advantage of the 
tridiagonal character of the Hamiltonian, as in ref. \cite{RSGF}.
Resonant tunneling at a nanotube heterostructure device has been shown to be robust and quite insensitive to 
both temperature and electron-phonon scattering at the temperature and voltage biases of interest \cite{Hyldgaard}.
Reference \cite{Hyldgaard} justified the use of an independent electron model for the calculation of transmission 
properties, which is adopted in the present approach.

Fig.1 shows examples of two mmm straight junctions: a (6,6)(12,0)(6,6) structure (fig.1b) and its 
complementary case (12,0)(6,6)(12,0) (fig.1c). The straight (6,6)(12,0) stable 
structure is formed by purely alternating pentagon-heptagon defects, which preserves a 6-fold 
rotational symmetry. This is true for any (n,n)(2n,0) junction, keeping n-fold rotational 
symmetry. As a result, we can see resonant peaks with maximum value of $2G_0$, and appearance of 
interference patterns of different character for the two straight cases (AZA and ZAZ, where A 
stands for (n,n) and Z for (2n,0)).
In addition to tunneling, bound states are confined in the structures, with
energies inside the range of the conducting bands. These states
are completely uncoupled from tunneling electrons, and do not
affect the conductance. They are manifested in the Green function as a
change of sign of the diagonal elements of $Re(G)$. In a non-purely alternating pentagon-hexagon defect
structure, such as a (6,6)(10,0) junction, rotational symmetry is greatly reduced, and thus electron
scattering between different angular momenta takes place, destroying the degeneracy. Our calculation
shows that this leads
to narrow resonant peaks with $1G_0$ conductance at their maximum (rather than $2G_0$ of fig.1b-c),
in a range from about 1 to 2 eV and from -1 to -2eV for a (6,6)(10,0)(6,6) (msm) and (10,0)(6,6)(10,0)
(sms).

{\it Simple Resonant Tunneling (SRT) and Anti Resonant Tunneling (ART) behaviours} -  
The remarkably regular interference pattern of the AZA structure (fig.1b) in the $-2.2 eV < E < 2.2 eV$ 
range suggests that the system behaves as a simple quantum interferometer 
(SRT behaviour). In such a case, 
the propagating waves in the interferometer can be described as a combination of a 'forward' and a 
'reflected' wave at each section of the structure: $\Psi = \psi^f+\psi^r$. Denoting the two junctions as $1$ and 
$2$, it is easy to show that the wave amplitude immediately at the right of $2$ is related to 
the amplitude immediately before $1$ by the expression \cite{Datta}
$\psi_{2+\delta}^f = {{t^2e^{i\phi}}\over {1-r^2e^{-2i\phi}}} \psi_{1-\delta}^f \equiv T\psi_{1-\delta}^f$, where 
$t$ and $r$ are the transmitivity and reflectivity of one of the two junctions alone ($t^2+r^2=1$). The 
denominator is a consequence of the multiple reflections between the two junctions of the 
interferometer (the derivation is no longer valid if the central part, between 1 and 2, is not a single 
chain). This implies that the transmission of the total 
system, $| T |^2$, is bound by two envelope curves which are solely functions of the conductance of the 
single (n,n)(2n,0) junction alone, as 
$1\geq |T|^2=|{{t^2e^{i\phi}}\over {1-e^{-2i\phi}(1-t^2)}}|^2 \geq ({t^2 \over {2-t^2}})^2$. Since we have 
two independent degenerate channels, the lower envelope curve for the total conductance of the 
AZA system, in units of $2e^2/h$, is 

\begin{eqnarray}
\sigma^{lim}_{AZA}=2({\sigma_{AZ} \over {4-\sigma_{AZ}}})^2
\end{eqnarray}

and the upper envelope is $\sigma_{AZA}=2$. And in fact this is indeed the case, as 
can be seen in fig.2(upper), where the 
conductance corresponding to states with angular momentum $L=\pm 1$ is plotted for different 
lengths of Z, and the minima are perfectly matched by expression (3).

  Since the envelopes of the interference pattern of a (n,n)(2n,0)(n,n) junction are fitted by an 
expression invoving just the conductance of the single (n,n)(2n,0) junction alone, one would 
wonder whether this would also be the case for the (2n,0)(n,n)(2n,0) junction. Fig.2(lower) shows that it 
is not. Despite the upper envelope is still a line at $2G_0$, the pattern now shows remarkable full 
antiresonances -zero conductance- (ART behaviour) in the resonant energy range, and no minima are matched by 
eq.(3). The difference between the AZA and ZAZ cases can be understood as follows.

One can transform the Hamiltonian basis from local orbitals ($\vert j \rangle$) to combinations for each layer 
(same z), on the case of the zigzag:

\begin{eqnarray}
\vert \gamma \rangle = {1\over{\sqrt{2n}}}\sum_j{e^{i\gamma j}\vert j \rangle}
\end{eqnarray}

and for the armchair:

\begin{eqnarray}
\vert \beta_{+(-)} \rangle = {{1(i)}\over{\sqrt{2n}}}\sum_j{e^{i\beta j}(\vert {j,1} \rangle \pm \vert {j,2} \rangle)}
\end{eqnarray}

where $\vert j,1(2) \rangle$ are the atomic orbitals at the two inequivalent locations of j, inside one ring
and p indexes the layer. The 
allowed values of the angular number are $\gamma =\pi L/n,  L=0,...,2n-1;  \beta=2\pi L/n,  L=0,...,n-1$.

In this new basis, the zigzag tube is reduced to a series of uncoupled independent chains with 
onsite energies $\epsilon=0$ and alternating hoppings $t_1=t$ and $t_2=2t \cos(\gamma/2)$. It is straightforward to check 
that this gives the proper bands of the zigzag tube. The armchair is reduced to double chains, 
rather than single, with onsite energies $\epsilon_+=-t, \epsilon_-=t$, 
and hoppings $t_{++}=-t \cos(\beta/2), t_{+-}=-t \sin(\beta/2), t_{--}= t \cos(\beta/2)$ (see fig.3).

Now, the two states $\vert \beta_+ \rangle$ and $\vert \beta_- \rangle$ on the (n,n) 
side couple only with two states,  $\vert \gamma = \beta/2 \rangle$  and $\vert \gamma' = \beta/2+\pi \rangle$
 , on the (2n,0) side. From this representation, it is 
apparent that a zigzag in the middle of the AZA behaves essentially as two single chains coupled 
to the double chain electrodes. The conductance of these two single chains alone (i.e. for a pure 
(2n,0) system) is shown as the dotted line in fig.2(upper). We see that for some energy ranges, only 
one of the chains' subband is present. For such range, we have a central region consisting of a 
single chain, and eq.(3) is thus valid. On the other hand, if the central region is consisting of a double 
chain, as in the ZAZ case, or the energy ranges where two chains are available for the Z in the 
AZA case, derivation of eq.(3) is no longer valid since the transmission of the AZ junction is no 
longer a scalar. In this case, the ZAZ system behaves as a single channel injecting electrons into 
two different channels, that afterwards merge again to drain in a single channel. When the energy 
is such that the electron arrives to the drain with opposite phase from each of the channels, an 
antiresonance takes place (ref. \cite{Anantram}).

Thus, the conditions for appearance of simple interferometer resonant pattern in the 
(n,n)(2n,0)(n,n) junctions can be stated: 1) it will only happen in the energy region such that the 
bands corresponding to $\beta_+$ and $\beta_-$ of the A region overlap with only one of the two subbands  
($\gamma=\beta/2$ or $\gamma=\beta/2-\pi$) of the Z region. 2) The opening of a new subband family with different $\beta$
leads to a superposition of two interference patterns, and destroys the simple pattern. In general, 
the new superimposed pattern has different periodicity than the other one, because each subband's 
minimum has different effective mass. This is different than in the case of a classical symmetric 
parabolic dispersion, where different $k_\parallel$ subbands have the same dispersion relation in $k_\perp$, and their 
resonant patterns superpose with the same periodicity in the interferometer. With nanotubes, on 
the contrary, the range of simple interference pattern is only until the opening of the new 
subband. This allows us to determine the interference energy ranges as a function of n.  In fig.4 we 
plot the lower edge of the subbands of the (n,n) system (negative slope curves) and the (2n,0) 
system (positive slope) as a function of n. The simple-interference region is limited by the two
thick solid lines,
fulfilling the two conditions stated above. We see that it goes from maximum to zero and 
viceversa cyclically, in periods of 6 units of n. in addition, the value of the maximum range, i.e. 
that at n=6, 12, 18, etc., decreases with n. Since its size is inversely proportional to the number of 
subbands in the system, it can be roughly approximated by ~3t/n. The analytic expression for the 
thick lines in fig.4 is given by the minimum for all q of
\begin{eqnarray}
Max\{\sin(\pi {q\over{n}}) , 1-2\cos({{\pi(q-n-j)}\over{2n}})\}, \nonumber\\
q=1,2,...,Int[n/2];    j=0,1.
\end{eqnarray}
where $j=0$ yields the lower line, and $j=1$ the upper one.

In conclusion, we have found occurence of antiresonances at general (2n,0)(n,n)(2n,0) nanotube
heterojunctions, and occurence of simple 'Fabry-Perot' type resonant tunneling at (n,n)(2n,0)(n,n) 
nanotube heterojunctions. The SRT conductance interference pattern has an upper constant envelope of 
$2G_0$, and a lower envelope that is fitted by $\sigma_{ABA}^{lim}=2(\sigma_{AB}/(4-\sigma_{AB}))^2$ 
in units of the quantum of conductance. This
simple pattern is due to the reduction of the (2n,0) part to a single channel at the energies of 
the interference. On the other hand, the reverse case of a (2n,0)(n,n)(2n,0) junction shows no 
longer a simple interference, but displays AntiResonant Tunneling. This is
because the (n,n) segment is reduced to a double chain and allows 
interference between two channels. The energy range of the simple interference pattern is  
within $0.5|t|<|E|<|t|$, and oscillates as a function of $n$, with 
a period of $6n$, being maximum at $n = 6, 
12, 18, ...$ and nearly 0 at $n = 9, 15, 21, ... $. The size of the maximum range becomes smaller 
with increasing $n$, roughly decreasing as $~3|t|/n$. Unlike parabolic dispersion band resonant 
interferometers, where resonator interference pattern takes place with the same period for all $k_\parallel$, the interference in nanotube ABA junctions
has different period for each subband, producing subsequent steps 
of superimposed periodicities when new subbands open. Possible application of resonant and anti-resonant
tunneling in nanotube based devices should take all these facts in consideration when the design of 
such nanostructures becomes experimentally available.

J. H. and N. M. acknowledge Prof. J.P.Lu for his support of the joint
research program between NASA Ames and University of North Carolina at
Chapel Hill.
\end{text}

\begin{figure}\begin{center} 
\mbox{\psfig{figure=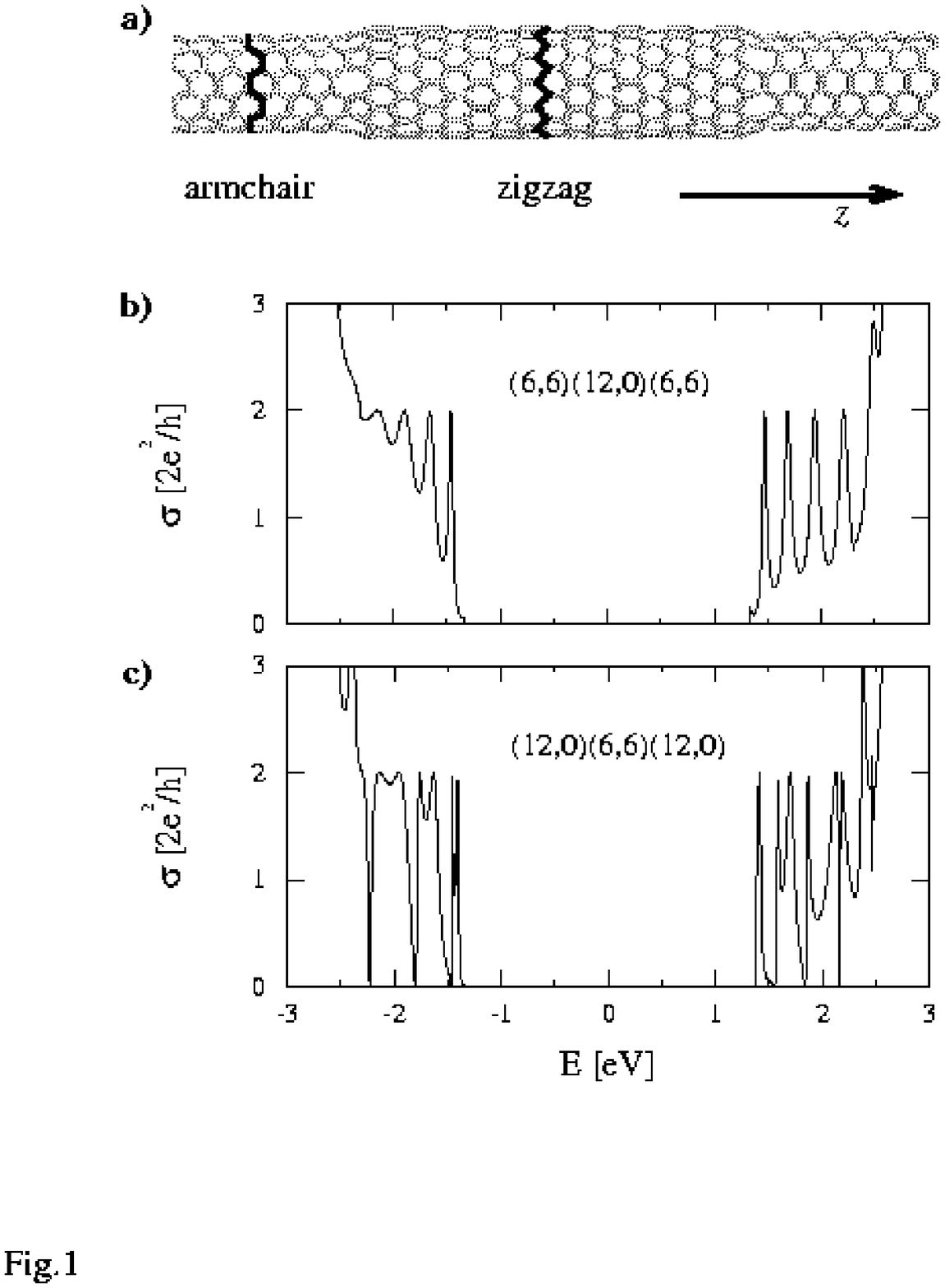,bbllx=95pt,bblly=110pt,bburx=525pt,bbury=665pt,width=10.25cm,angle=0,clip=}}
\caption{Geometry and conductance vs. energy in two examples of straight ABA type nanotube 
heterostructures. a) Geometrical structure of a (6,6)(12,0)(6,6) junction; 
b) conductance of a (6,6)(12,0)(6,6) m-m-m, with 8 unit cells in the (12,0) section; 
c) conductance of (12,0)(6,6)(12,0) m-m-m, with 13 unit cells in the (6,6) section. 
See explanations in main text.}

\label{fig1} 
\end{center}
\end{figure} 

\begin{figure}\begin{center} 
\mbox{\psfig{figure=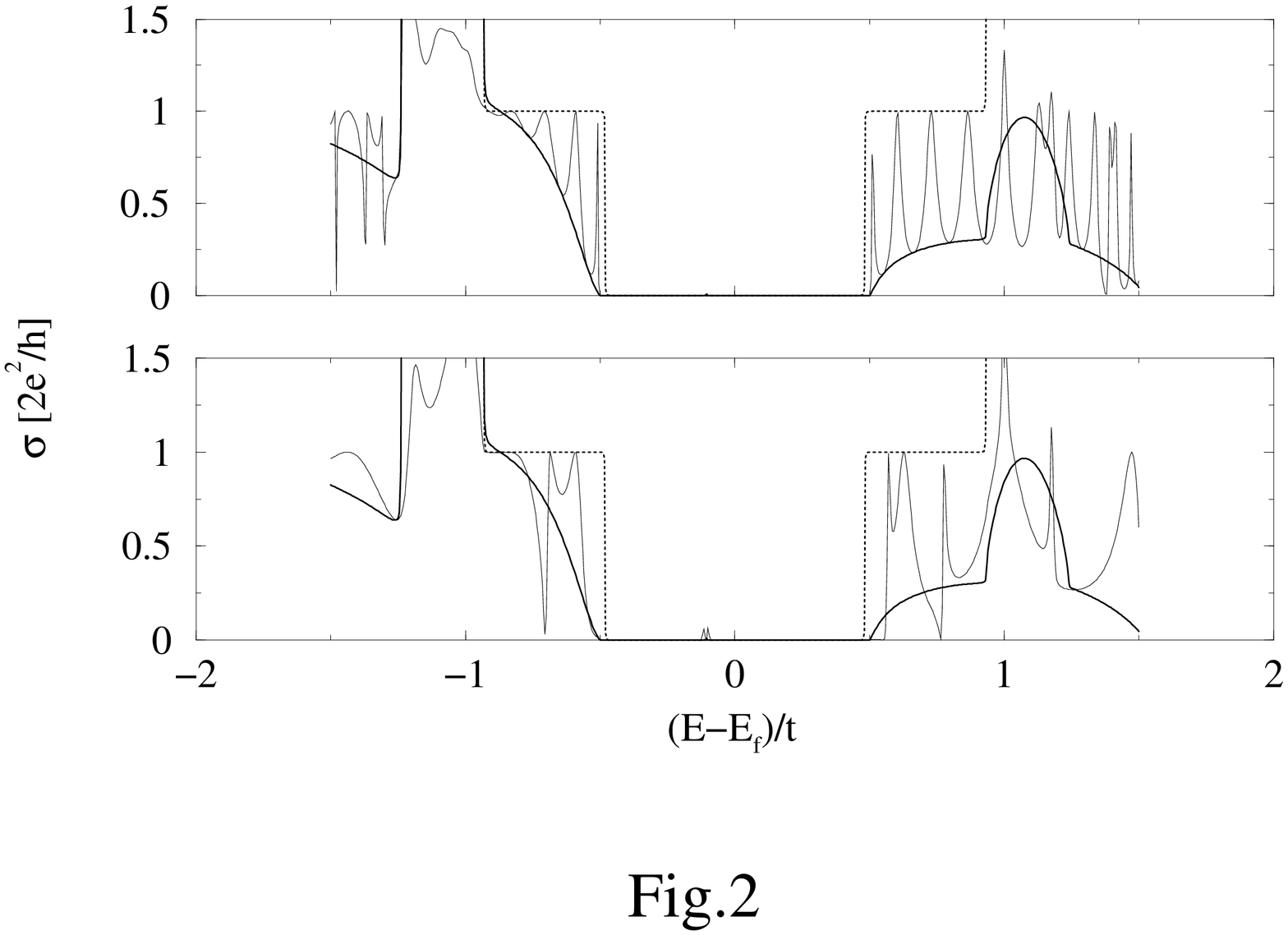,bbllx=26pt,bblly=58pt,bburx=493pt,bbury=723pt,width=12.25cm,angle=-90,clip=}}
\caption{{\it{Upper}}: SRT of electrons with $L=\pm1$ in a (n,n)(2n,0)(n,n) heterostructure.
The case shown corresponds to n=6, and 5 unit cells in the (2n,0) part.
The dotted line above the curve
corresponds to the conductance of the two channels with $L=\pm1$ of an infinitely long (2n,0) tube.
When only one of these channels is available, the resonant minima are perfectly matched 
by $\sigma^{lim}_{ABA}=2(\sigma_{AB}/(4-\sigma_{AB}))^2$ (thick solid line), 
which is a solely function of the conductance of a single AB junction.
When the two channels are available (dotted line goes to 2) the simple interference pattern is destroyed,
and antiresonances appear.
{\it{Lower}}: ART of electrons with $L=\pm1$ in a (2n,0)(n,n)(2n,0) 
heterostructure (reverse of case in fig.2a), for 5 unit cells in the (n,n) part. 
The (n,n) section has two channels available at all energies,
which implies that full antiresonances occur. The curve given by eq.(3) is also plotted, to show
that it does not fit any of the minima in this case.}
\label{fig2} 
\end{center}
\end{figure} 

\begin{figure}\begin{center} 
\mbox{\psfig{figure=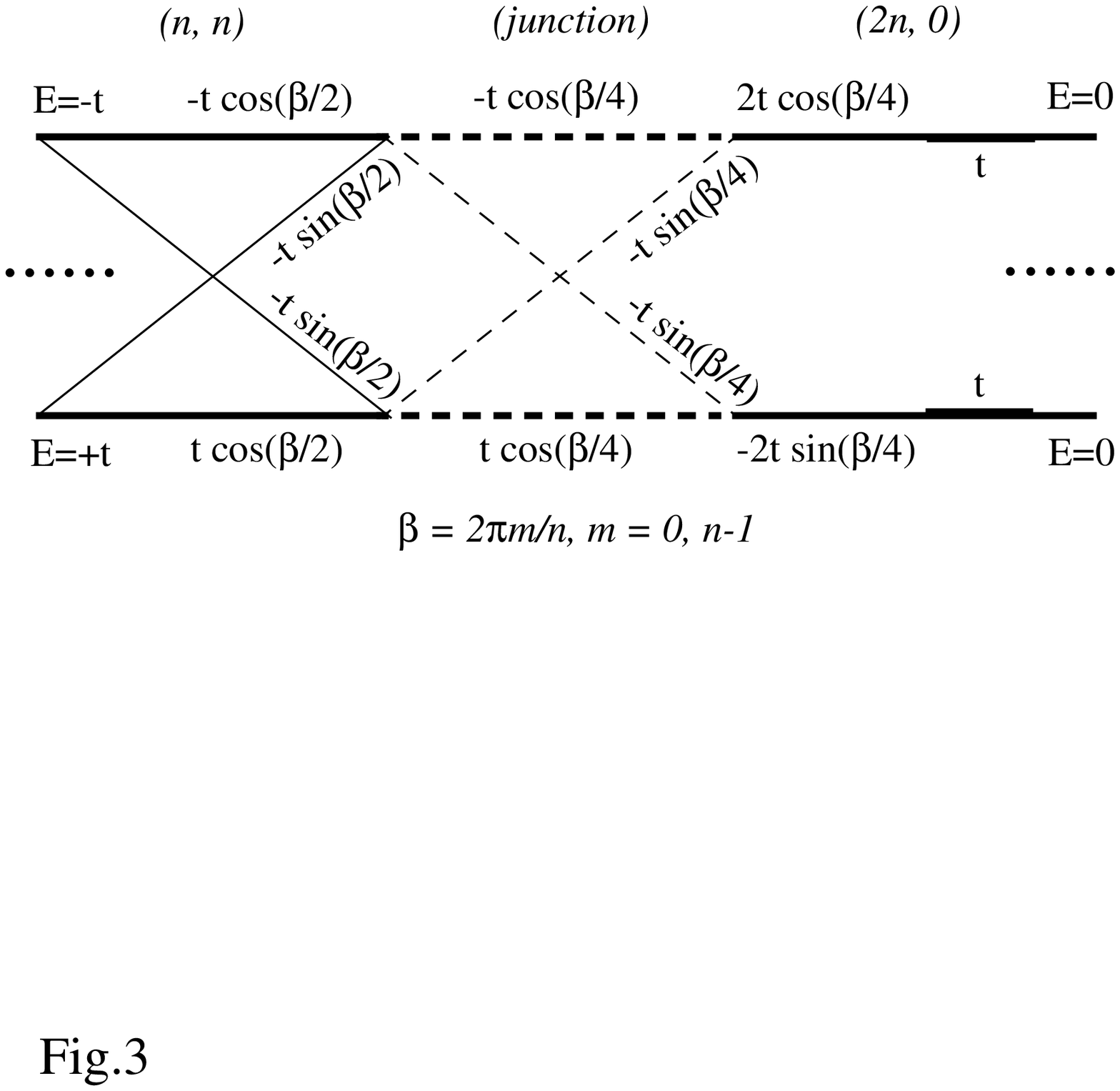,bbllx=31pt,bblly=299pt,bburx=589pt,bbury=639pt,width=12.25cm,clip=}}
\caption{Uncoupled double chains to which the system is reduced by transformation (4,5). The 
onsite energies and hoppings are shown.}
\label{fig3} 
\end{center}
\end{figure} 

\begin{figure}\begin{center} 
\mbox{\psfig{figure=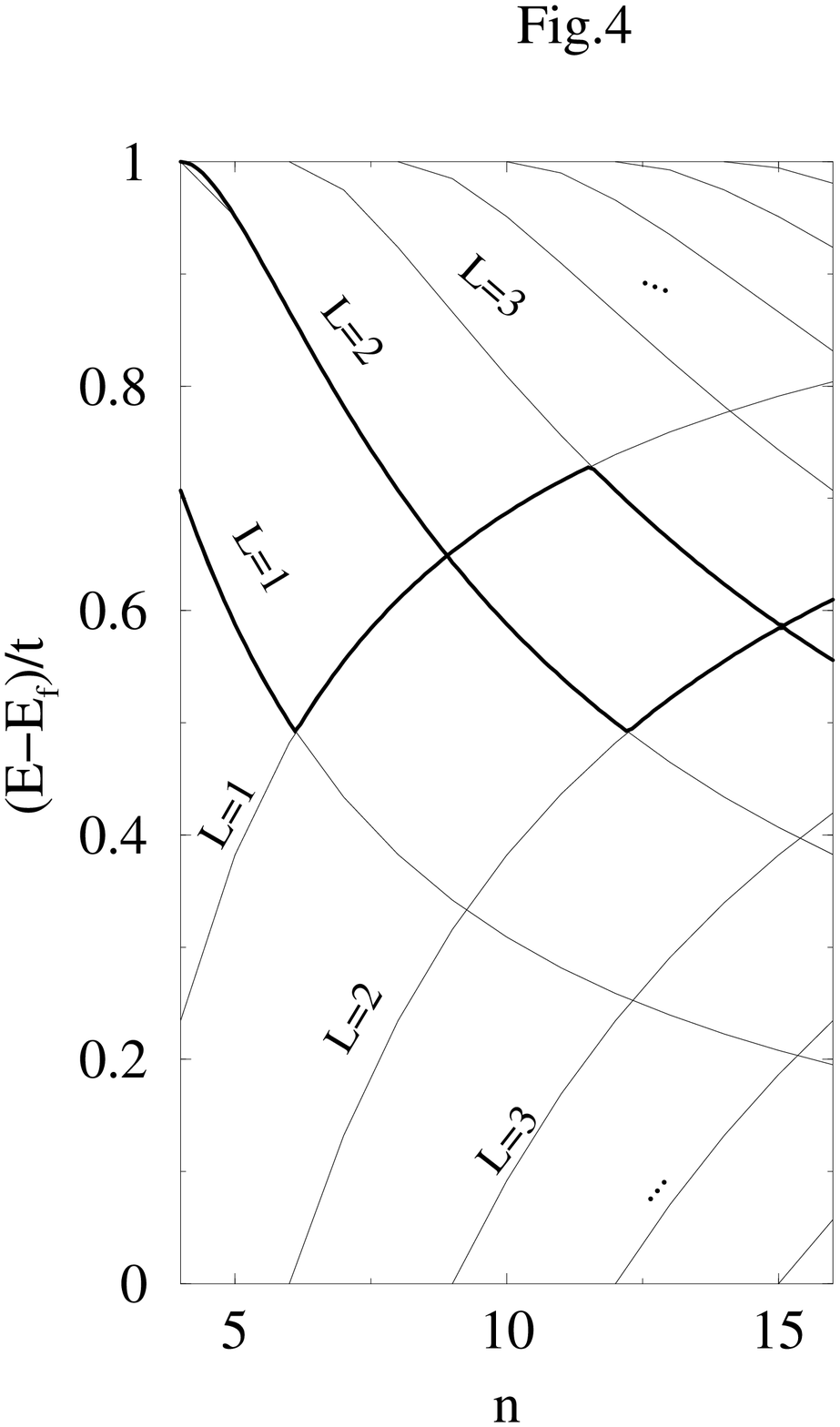,bbllx=10pt,bblly=60pt,bburx=450pt,bbury=700pt,width=7.25cm,clip=}}
\caption{Energy range in which SRT occurs, as a function of n, for (n,n)(2n,0)(n,n) type 
heterostructures, is delimited by the thick solid lines. 
The negative slope curves correspond to the lower edges of the (n,n) section
subbands, and the positive slope curves mark the lower edges of the (2n,0) section subbands
(not all these are plotted, to preserve the graph's clarity). 
In order to obtain the biggest possible range of 
SRT, one should choose $n$ being a multiple of 6, and within these, the smallest possible n. On the
other hand, systems with $n=6j+3, j=1,2,...$ will display no SRT range.}
\label{fig4} 
\end{center}
\end{figure}


\begin{references}
\bibitem[*]{byline} To whom correspondence should be addressed.
\bibitem{bend} Bend junctions can exhibit resonant patterns around the Fermi level in some cases. The conditions for this to occur will be published elsewhere.
\end{references}
\end{document}